\documentclass[reprint]{revtex4-1}
\usepackage{amsmath, amsfonts, amsthm, amssymb,epsfig} 
\usepackage{graphicx}
\usepackage{stfloats}
\usepackage{epstopdf}
\usepackage{hyphenat}
\usepackage{dcolumn}
\usepackage{bm}

\begin{document}

\title{Fast and accurate shot noise measurements on atomic-size junctions in the MHz regime}

\author{Sumit Tewari}
\affiliation{Huygens-Kamerlingh Onnes Laboratorium, Universiteit Leiden, Postbus 9504, 2300 Leiden, The Netherlands.}
\author{Carlos Sabater}
\altaffiliation{Present address: Chemical Physics Department, Weizmann Institute of Science, 76100 Rehovot, Israel.}
\affiliation{Huygens-Kamerlingh Onnes Laboratorium, Universiteit Leiden, Postbus 9504, 2300 Leiden, The Netherlands.}
\author{Manohar Kumar}
\altaffiliation{Present address: Laboratoire Pierre Aigrain - ENS 24, rue Lhomond, 75231 Paris Cedex 05, France.}
\affiliation{Huygens-Kamerlingh Onnes Laboratorium, Universiteit Leiden, Postbus 9504, 2300 Leiden, The Netherlands.}
\author{Stefan Stahl}
\affiliation{Stahl Electronics, Hauptstrasse 15, 67582 Mettenhein, Germany}
\author{Bert Crama}
\affiliation{Huygens-Kamerlingh Onnes Laboratorium, Universiteit Leiden, Postbus 9504, 2300 Leiden, The Netherlands.}
\author{Jan M. van Ruitenbeek}
\affiliation{Huygens-Kamerlingh Onnes Laboratorium, Universiteit Leiden, Postbus 9504, 2300 Leiden, The Netherlands.}
 \email{ruitenbeek@physics.leidenuniv.nl}

\begin{abstract}
Shot noise measurements on atomic and molecular junctions provide rich information about the quantum transport properties of the junctions and on the inelastic scattering events taking place in the process. Dissipation at the nanoscale, a problem of central interest in nano-electronics, can be studied in its most explicit and simplified form. Here, we describe a measurement technique that permits extending previous noise measurements to a much higher frequency range, and to much higher bias voltage range, while maintaining a high accuracy in noise and conductance. We also demonstrate the advantages of having access to the spectral information for diagnostics.
\end{abstract}

\maketitle

\section*{INTRODUCTION}
Atomic-size contacts and single-molecule junctions have proven to be wonderful platforms for exploring the properties of electron transport in the quantum conductance regime.\cite{agrait03,cuevas10} The structure of quantized conductance channels and their connection to atomic and molecular orbitals has been elucidated. \cite{scheer97,scheer98,smit02,kiguchi08a,berndt10}  For such investigations mere measurement of the conductance of the junction provides insufficient information on the composition of the junction. Indeed, in the coherent quantum limit the conductance is given by Landauer's expression as $G=G_{0}\sum T_n$, where $G_{0}=2e^2/h$ is the conductance quantum, and the sum runs over the transmission probabilities $T_n$ of each of the participating conductance channels. This sum hides all the details of how many channels are involved and what is the relative weight of each of these. 

Several techniques have been developed over the years that give access to more detailed knowledge. The most powerful of these methods exploits multiple Andreev reflection, using superconducting metal leads.\cite{scheer97,scheer98} The combination of the quantum structure of the atomic-size junction with the energy gap structure of the superconducting leads gives rise to strongly non-linear current-voltage (IV) curves with rich finestructure, from which in many cases the full set of transmission probabilities $\{T_{n}\}$ can be extracted. This set $\{T_{n}\}$ is often referred to as the mesoscopic PIN code of the junction, since knowledge of this code is sufficient to predict quantitatively all other electronic properties of the junction.\cite{cron01,levy01}

Shot noise has become an equally important tool in analyzing the quantum conductance structure of atomic-size junctions. The information that it provides is more restricted compared to the use of superconducting subgap structure, but it has the advantage that it is more widely applicable, since many junctions of interest cannot easily be formed by the use of superconducting leads. Moreover, the requirement of cooling to well below the superconducting critical temperature does not apply, by which the method again gains wider applicability.

The study of shot noise in atomic-size junctions has gained further interest by the observation of signals due to inelastic scattering of the electrons.\cite{kumar12,chen15} Atomic and molecular junctions offer a unique opportunity for observing inelastic contributions to noise in detail, and the effect opens new avenues for investigating atomic and molecular junctions.\cite{avriller09,haubt09,schmidt09}

Shot noise results from the discrete character of the charge of electrons. For a tunnel junction the transmission of electrons is a stochastic process described by a Poissonian distribution, giving rise to a current noise power spectral density $S_I = 2 e I$, where $e$ is the electron charge and $I$ is the mean current.\cite{schottky1918} A quantum conductance channel with perfect transmission has zero shot noise because the electron wavefunction is completely delocalized between left and right contacts, so that the current loses its classical particle-like discreteness. A general expression for noise at finite temperature and voltage, and for an arbitrary set $\{T_{n}\}$ has been obtained by Lesovik and others\cite{lesovik89,buttiker90,blanter00}, and can be represented as,\cite{kumar12}
\begin{equation}
Y(V,T) = F\, \left[ X(V,T) -1 \right]. \label{eq.YvsX}
\end{equation}
Here, $Y(V,T) = (S(V,T) - S(0,T) ) / S(0,T) $ is the reduced noise power.\footnote{We discuss voltage noise rather than current noise, because this is the parameter we directly measure. The conversion is straightforward, obtained through the resistance $R_{\rm s}$ of the junction as $S_V = S_I R_{\rm s}^2$. } It measures the noise above the zero-voltage thermal noise, and scaled to the thermal noise. The control parameter $X(V,T)$ is given by  $X(V,T) = x \coth x$, where $x$ is the ratio of the voltage and temperature in the experiment $x = eV/2k_{\rm B}T$. The noise spectrum is white up to very high frequencies $\omega$ for which $\hbar\omega$ becomes comparable to $eV$ or $k_{\rm B}T$.
The so-called Fano factor $F$ gives the slope of the reduced noise power as a function of $X$, and this is the quantity containing the relevant information, which we will be interested in measuring. In the absence of inelastic scattering $F$ depends only on the transmission probabilities $\{T_{n}\}$, as
\begin{equation}
F = \frac{\sum T_n (1-T_n)}{\sum T_n}. \label{eq.Fano-factor}
\end{equation}

The central problem of concern in noise measurements is the impedance of the device under test, in this case the junction resistance $R_{\rm s}$. The scale is set by the quantum of resistance $R_{\rm q} = 1/G_{0}$, which is about 12.9~k$\Omega$. When the junction contains many channels with high transmission probabilities the information obtained from the Fano factor $F$ in combination with the conductance $G$ is very limited, because there are too many unknown parameters. For this reason we will be mostly interested in the range of resistances $ R_{\rm s} \gtrsim 1 \mathrm{~k}\Omega$. On the other hand, when the junction resistance $R_{\rm s}$ becomes very high this fact alone implies that all conductance channels have a small transmission probabilitity. 
From the expression (\ref{eq.Fano-factor}) we conclude that the Fano factor approaches 1 in this limit, and it will be difficult to draw strong conclusions from small deviations from unity. In practice this limits the range of interest to 1~k$\Omega \lesssim  R_{\rm s} \lesssim 100$~k$\Omega $, and $R_{\rm s} \simeq 10$~k$\Omega$ being of particular interest.

Since shot noise is white up to very high frequencies we may choose to measure it in any convenient frequency window. However, practical limitations are set from two sides. At the low side of the spectrum ubiquitous 1/f-noise dominates the measured noise. This 1/f-noise is attributed to two-level fluctuations of various origin,\cite{kogan96} and is present in the amplifiers as well as in the sample itself. The latter source of noise is particularly troubling since it is resistance noise, for which the voltage noise power increases with the applied voltage bias over the junction as $V^2$. Since shot noise  is linear in $V$, for high bias 1/f noise is likely to dominate. Since 1/f-noise decays roughly as $\omega^{-1}$ we will want to measure the noise spectrum at high frequencies.

The high end of the spectrum is limited by the bandwidth of the measurement, which is set by the $RC$ time of the combination of the junction resistance $R_{\rm s}$ together with the capacitance of the cables and the amplifier input. With a typical cable capacitance of 100~pF/m, two cables connected, and $R_{\rm s}=10$~k$\Omega$ the bandwidth is limited by a corner frequency of $f_{\rm c} = 1/2\pi R C \simeq 80$~kHz. For metal atomic junctions at moderate voltage bias this leaves a window of about two decades between 1 and 100~kHz for measurement.\cite{brom99}

Let us now define the goals of the present experimental set-up: (1) we want to increase the voltage bias range in order to explore inelastic effects at higher bias, which will dramatically increase the amplitude of 1/f-noise and push the lower limit of the useful spectral range to higher frequencies. This is further exacerbated by moving from metallic atomic junctions to organic molecular junctions, which display a higher tendency of developing two-level fluctuations. In order to compensate for the loss in spectral range we want to push the upper limit of the spectrum to higher frequencies.
(2) Noise measurements at low frequencies are slow. The measurement time is set by the intrinsic stochastic character and the associated requirement of long averaging times needed for determining accurate noise power levels. We will be interested in high accuracy and speed, such that we can detect small inelastic corrections and that we can follow these quasi-continuously as a function of control parameters such as bias voltage or mechanical stretching of the junction. The recording time for each data acquisition scales inversely with the lowest frequency of the spectral window. By shifting the lower limit of the frequency window from 1~kHz to  100~kHz the speed of measurement is increased hundred fold, provided the speed of data manipulation can be adjusted to the high data acquisition rates.

Other factors of concern are the noise introduced by the electronics of the amplifier chain, and pick up of spurious signals. Spurious signals are easily identified when the Fourier spectrum over a wide range is available for inspection, which is why we have decided to focus on broad band spectrum-based noise measurements. Amplifier noise may limit the accuracy, which is why we have decided to adopt the method introduced by Glattli and coworkers,\cite{kumar96} based on measuring the noise signal twice, through two parallel and independent chains of amplifiers with high input impedance. The noise signal is then obtained from calculating the cross spectrum of the two signals, from which the uncorrelated noise of the two amplifiers is strongly suppressed after averaging. In order to widen the bandwidth the first-stage amplifiers are mounted in the cryogenic environment at close proximity to the break junction device. This allows us to push  the upper limit of the usable frequency spectral range up by nearly two orders of magnitude, from 100~kHz used previously\cite{brom99,cron01,vardimon13} to 6~MHz, similar to the approach used by DiCarlo {\it et al.} \cite{dicarlo06} and by Jullien \cite{jullien15}  for measuring the cross correlation of two separate sample signals.

Several alternative strategies have been explored by other groups, depending on the problem to be addressed. Thibault {\it et al.}\cite{thibault15} demonstrated noise measurements over a very wide frequency range, even into the GHz range. The fact that they could tune the resistance of their sample to match the 50~$\Omega$ impedance of the cables and amplifier input explains why this was possible in their experiments. In general, the applicability of the technique is subject to limitations set by the actual resistance of the sample under test. Yet, even for high sample resistances it has been demonstrated  that noise can be effectively measured in the GHz regime.\cite{reznikov95,wheeler10,parmentier11}  
Although there are advantages of noise detection in the GHz regime, the main disadvantage is that the Fourier spectrum is not readily available for diagnostics and detection of spurious signals. In these experiments only the integrated noise is detected by a power detector. 

Having decided that we want to employ high-input impedance amplifiers, as a consequence the bandwidth is limited to a few MHz, resulting from the sample resistance in the range of the quantum resistance and a few pF of input capacitance that cannot be avoided. Two further strategies have been demonstrated for pushing the bandwidth still higher. Birk {\it et al.} employed active feedback of the measured AC signal, back to the input amplifier stage, in order to compensate for the input capacitance.\cite{birk96} This allows increasing the bandwidth by two to three orders of magnitude. Although this will be a very attractive option when we want to explore higher junction resistances, at this point it is not needed. It would add amplifier noise and would not be compatible with the configuration of two parallel sets of amplifiers. A second strategy, which was also implemented by DiCarlo and coworkers\cite{dicarlo06} and by Jullien \cite{jullien15} is to add an inductor in parallel to the capacitance in order to convert it from a low-pass filter into a band pass filter centered at higher frequencies. Although this reduces the bandwidth to only about 100kHz, this is still a very interesting option for further raising the measurement frequencies. 

\section*{Principles of the method}

Atomic-size contacts and molecular junctions are formed by mechanical adjustment of the gap between two metal electrodes using the mechanically controllable break junction (MCBJ) technique.\cite{muller92,ruitenbeek96} This allows forming fresh metal electrodes by breaking of a weak spot produced in a wire or microfabricated bridge. The mechanical adjustment is controlled by a combination of a mechanical screw translation stage and a piezo electric element for fine adjustments. The junction can be broken at liquid-He temperatures under vacuum, so that contamination can be avoided for extended periods of time. Molecules can be introduced to the junction, either through a capillary connected to a room-temperature vapor source, or by deposition from solution prior to cool-down.  A discussion of the benefits of this method has been presented elsewhere,\cite{agrait03} and details of its present implementation adjusted for high-frequency noise measurements will be given below.

Low temperatures are required for suppressing atom mobility, for spectral resolution of vibration modes of the junction in differential conductance measurements, and for reducing 1/f-noise and thermal noise of the junction. Since variation of temperature is usually not required, a practical choice is given by the temperature of a liquid He bath at ambient pressure, about 4.2~K. This temperature is low enough such as to make the sample thermal noise comparable to instrumental noise sources.  The mechanical construction of the vacuum can takes the form of a dip-stick that can be lowered directly into a liquid-He storage vessel, which permits fast cycling between experiments. 

Wide-band noise measurements on the junctions are permitted by mounting two parallel low-noise cryogenic amplifiers at close proximity to the junction, by which cable capacitance is limited. These first-stage amplifiers have a high input impedance, and a 75~$\Omega$ output. The junction is biased from a voltage source through a series resistor, and the conductance of the junction is measured by means of an audio-frequency lock-in technique. Decoupling of the cable capacitances of the biasing and the voltage sensing line is accomplished by inserting a second series resistor near the junction in the cold. The frequency window chosen for the noise analysis is 1 to 6~MHz. While the transfer characteristics of the electronic circuit in this window is already decaying as a result of the unavoidable input capacitance, the transfer characteristics can be accurately taken into account and the signal-to-noise permits measuring Fano factors with an accuracy better than 1\%. 

The higher spectral frequency range permits shorter acquisition times, and for the data analysis to be able to keep up with acquisition the Fourier transforms, computation of cross spectra, and averaging are handled by a Field Programmable Gate Array (FPGA) unit. All other processing is done under LabView control on a standard desktop computer. Table~I summarizes the properties of the system, and a detailed description of each of the components follows below.

\begin{center}
TABLE I. Parameters characterizing the instrument.\label{table}
\begin{tabular}{| l | c |}
\hline
Spectral acquisition speed* & 12 s$^{-1}$  \\
Noise spectral frequency range & 0.1 -- 10~MHz\\
Sample conductance range & 0.1 -- 10~$G_0$  \\ 
Bias voltage range &  $\pm$~100~mV\\
Accuracy of conductance measurement & $< 1$\% \\
Accuracy of Fano factor & $< 0.01$ \\
Temperature &  $4.4\pm0.2$~K\\
\hline
\end{tabular}\\
* For  a standard number of 10,000 times averaging.
\end{center}

\section*{Description of the instrument}

\subsection*{Mechanical design}
\begin{figure}[!h]
\includegraphics[scale =0.32]{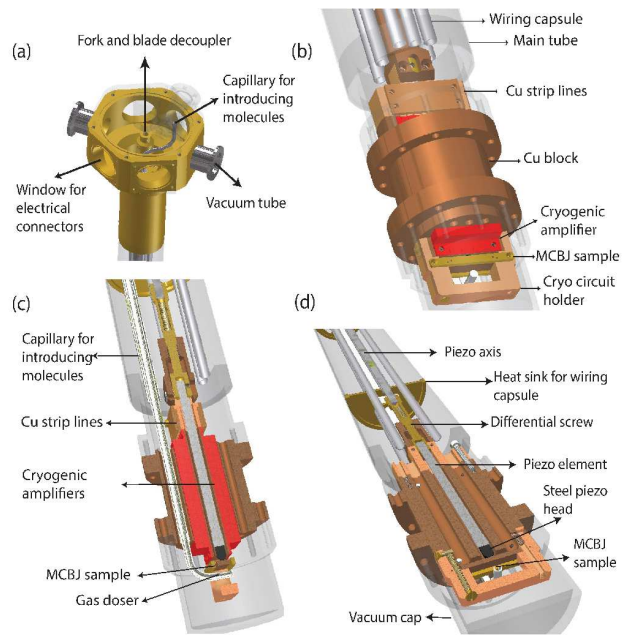}
\caption {(Color online) Lay out of the dipstick. (a) Top flange with connector head. (b) Cu block serving for thermalization of the cryo-electronics. (c) Cross section of the lower part of the insert showing the position of the amplifiers (shown schematically in red) inside the Cu block . (d) Cross section of the lower part of the insert showing the position of the piezo element and the mounting of the sample. \label{dipstick}}
\end{figure}

The dipstick has been designed to fit directly into a He transport vessel, which sets its dimensions as 50~mm in diameter and 1.5~m long. At the top of the tube a connector head is mounted, shown with some of the covers removed  in Fig.~\ref{dipstick}a. It has ports for electrical connections, vacuum pumps and pressure gauges, a connection to a capillary for injection of molecular vapor, and a rotary feed-through in the top lid for mechanical adjustment of the junction. The central tube contains ducts for wiring, for the gas dosing capillary, and an axis for mechanical control.
The dipstick unit accommodates the first-stage noise amplifiers at close proximity of the break junction. The working temperature of the amplifiers affects gain and noise performances so that it is of importance to take care of proper thermalization of the amplifiers and of the all the wires going to and out from the cryogenic amplifiers. The thermalization is provided by a copper block (Fig.~\ref{dipstick}b) that connects the main tube to the sample space, houses the amplifiers (Fig.~\ref{dipstick}c), and is in direct contact with the helium bath. Copper wool is added between the gold plated base of the low temperature amplifiers and the copper bock for better thermalization.  The capillary tube for admitting gasses is isolated from the Cu block using a Teflon ring and has an extension at the cryo end with a gas dozer nozzle facing the sample.

The wires running down the central tube are held by capsules, which are formed by semi-cylindrical tubes (a tube cut in half along its length). All the wires coming from the head at room temperature are Manganin (Cu($86\%$) Mn($12\%$) Ni($2\%$)) wires, which provide a good compromise between a high electrical conductance and a low thermal conductance. The wires are glued onto the inside of the semi-cylindrical tubes by GE varnish.  These half-cylindrical tubes are inserted by sliding from the top into the fine hollow tubes, which in turn are fixed to the main tube of the dipstick using silver welding for good thermal and mechanical anchoring. This arrangement has the advantage that it helps anchoring the wires thermally and mechanically to the inner wall of the rod efficiently, while the wires can be easily removed from the wiring capsule for maintenance and modifications. At the bottom end all wires are soldered onto electrically isolated Cu strips on the cold finger extending from the Cu block, where they are cooled efficiently by the liquid He bath. The axis for the mechanical rotation drive is also coupled to the copper block for thermalization. 

A mounting stage for the sample is fixed to the bottom of the copper block. The sample has the form of a bendable substrate of about 3 by 20 mm carrying the notched sample wire. The side carrying the notched sample wire faces down and is held against two fixed supports at either end. The top of the piezo element pushes in the middle against the back of the substrate, from above, for control of the bending. The sample space is enclosed by a brass cap that is mounted onto the lower end of the copper block by ten 3~mm 
bolts, and sealed with a thin indium ring. The sample space is in contact with the main volume of the central rod through openings in the copper block, permitting efficient vacuum pumping. During the measurements the actual pressure in the sample space is further improved by cryo-pumping of the walls of the container.  A container with active coal is mounted in the cryo cap for enhanced cryogenic pumping. Changing samples just requires opening the cap, so that only the input side of the low temperature amplifiers and the sample are exposed. All the rest of the wiring is fixed within the copper block.
By placing the copper block directly in contact with the He bath an efficient thermalization of the sample and the amplifiers is achieved, giving a cooling time of about $2$hrs to the base temperature of $4.2$K.

We paid special attention to attaining good mechanical stability, since future measurement runs stretching over several hours of measurement time on a single contact setting are anticipated. Instabilities in the atomic contacts can be grouped into two classes, those inherent to the properties of the sample, \textit{i.e.} due to the microscopic structure and atomic arrangement very close to the atomic contact, and the other due to external electro-mechanical disturbances to the atomic contact. Instabilities inherent to the sample can often be improved by annealing the sample in the cryogenic vacuum at high currents. External mechanical instabilities can result from electrical noise in the piezo voltage, or vibrations coupling mechanically or acoustically into the setup. 
The piezo wires are held apart from the sensitive measurement wires by mounting them in separate wire tubes. For mechanical stabilization the screw drive axis can be decoupled from the lower parts using a fork blade configuration, which is positioned at the top of the insert. Further reduction of mechanical and acoustic disturbance is achieved in the usual way by placing the experiment inside an acoustically shielded Faraday cage suspended on pneumatic vibration dampers. 

\subsection*{Analog electronics}

\subsubsection*{Low-frequency components}

\begin{figure}[h!]
\includegraphics[scale =0.37]{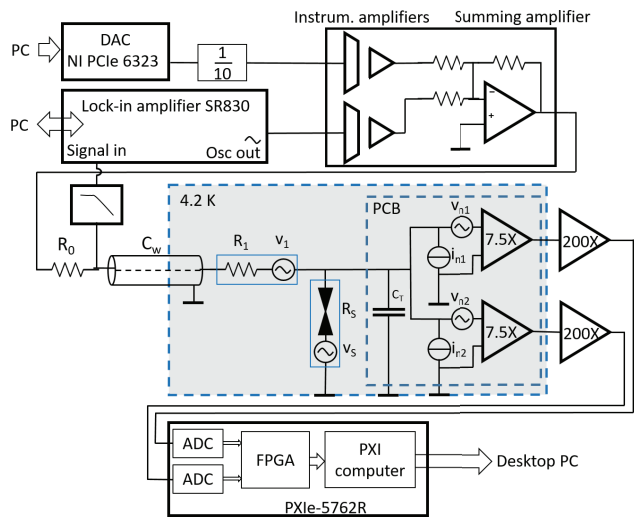}
\caption{Schematic of the circuit used for measurements of conductance and noise on break junction samples (indicated by the two touching black triangles). The components enclosed in the shaded block are at liquid-helium temperatures. The noise sources of the sample, the resistor $R_1$ and the current noise and voltage noise sources at the inputs of the amplifiers have been given explicit symbols.  \label{fig:circuit}}
\end{figure}

Figure~\ref{fig:circuit} shows the lay out of the electronic circuit. A central star grounding scheme for all signal circuits was used with the star point at the cryo head, in order to avoid ground loops.  Apart from the two parallel noise signal amplifiers the atomic-size junction is connected through a single line that provides the voltage bias and permits measurement of the conductance of the junction. Each additional line connected to the sample adds capacitance and reduces the bandwidth. Therefore, the number of lines was limited to just one, and a resistor $R_1$ near the sample serves to decouple the capacitance of the wire from the sample noise signal. The choice of this resistor $R_1 = 10 {\rm ~k}\Omega$ is a compromise between optimal decoupling, for which the resistance should be as high as possible, the desired accuracy in the conductance measurement, and Joule heating in the resistor at high bias. An inductor has been considered as an alternative, but inductors have the disadvantages that magnetic cores cause non-linearities at high dc bias currents, and unfilled coils require too much space. Since the frequency response and the thermal noise associated with $R_1$ are known they can be included in the signal analysis described below. 

The differential conductance as a function of applied dc bias is measured employing a two-probe ac technique by applying a small-amplitude modulation voltage on the bias line, corresponding to values of a few~mV$_{\rm rms}$ across the sample, at a frequency of 677~Hz. The modulation is supplied by the internal oscillator of a SR830 lock-in amplifier, and the resulting voltage at the top of the insert is measured by the same instrument. We found that connecting the input of the lock-in introduces high-amplitude disturbance over the full range of our noise spectra, but this can be fully suppressed by means of a low-pass filter, for which we employ a T-filter configuration having 1~M$\Omega$ at the signal side, 100pF to ground, followed by 10~k$\Omega$ towards the input of the lock-in. The resistance of the junction is calculated back from the voltage division over the biasing resistor $R_0$  (100~k$\Omega$) and the sample resistance in series with the resistor $R_1$ (10~k$\Omega$), taking phase shifts and signal reduction due to known (stray) capacitances into account. There is some loss in accuracy compared to a scheme with a separate voltage sensing line at the sample, behind $R_1$. However, additional lines introduce additional capacitance which deteriorate the bandwidth. We have also tested a low-temperature buffer amplifier for dc or audio-frequency voltage sensing, but we were not satisfied with the accuracy and reproducibility after cycling the amplifiers to room temperature. We further tested a miniature relay near the junction, but found that the switching caused unwanted junction breaks, both, as result of mechanical vibrations as well as electrical spikes. The solution given here is simple and sufficiently accurate for our purposes. 
 The overall accuracy of the ac conductance measurement was tested against fixed calibration resistors, giving reproducibility better than 1\%.

The dc bias voltage is supplied directly from a National Instruments data acquisition card (PCI DAQ-NI 6221). The DAQ card offers a digital-to-analog convertor with 16 bit digital accuracy, and is  controled from the desktop computer through a dedicated Labview routine. We were initially suspecting that taking the bias voltage directly from the DAQ would introduce spurious signals in the spectra, but we found that the spectrum remains clean, even without any additional filtering of the bias source. However, in order to break ground loops we have introduced instrumentation amplifiers in the bias line, and the adder that combines the ac and dc bias signals is powered from a battery pack. 

The combination of wires running down the insert bridging the temperature gradient between room temperature and liquid-He temperatures gives rise to a thermal voltage of 0.56~mV. In our circuit nearly 90\% of this dc thermovoltage drops over $R_0$, so that the influence on the dc bias of the sample is very small. By improving the wiring symmetry the thermovoltage can be further reduced, but in our case we take this small offset into account in the numerical analysis of the data.

The fine tuning of the contact size at the atomic scale is controlled by a piezo element, as shown in Fig.~\ref{dipstick}d. The potential on the piezo element is controlled using a digital-to-analog channel on the same DAQ card, through a Physik Instrumente E-665 high-voltage amplifier.

The temperature of the junction is measured by a standard ruthenium oxide resistance thermometer in a four-probe low-frequency ac resistance bridge. The resistance thermometer is glued to the brass break junction support at the point where the sample substrate is clamped. It is mostly used to verify deviations from the standard operation temperature of 4.2K, and this point has been accurately determined by placing the thermometer in contact with the liquid helium.

\subsubsection*{High-frequency low-noise amplifiers}
The cryogenic amplifiers are designed in close collaboration with Stahl Electronics. The design constraints for the amplifiers are given by the desired input current and voltage noise levels, by requirements for effective thermalization of all active and passive elements on the printed circuit board, and by size limitations.  The design of the cryogenic amplifiers was insipred by the low noise amplifiers described by A. T.-J. Lee.\cite{lee89} The amplifiers have two stages, the first of which is a dual-channel low noise cryogenic amplifier with operational frequency window of 160~kHz to 50~MHz.  The output of this stage, with an impedance of 75~$\Omega$ is coupled to the input of the second-stage low noise amplifiers at room temperature, with operational frequency window of 400~Hz to 140~MHz, for which we have selected the NF SA-230F5. The advantage of this separation is the reduced heat load on He bath and the smaller size of the  cryogenic amplifier. For the cryogenic stage we have used high electron mobility field-effect transistors (HEMT-FETs,  ATF34143). The heat load on the He bath is limited to $12$~mW. The cryogenic amplifiers as well as the room-temperature amplifiers are powered from battery packs.

\begin{figure}[!h]
\includegraphics[scale =0.34]{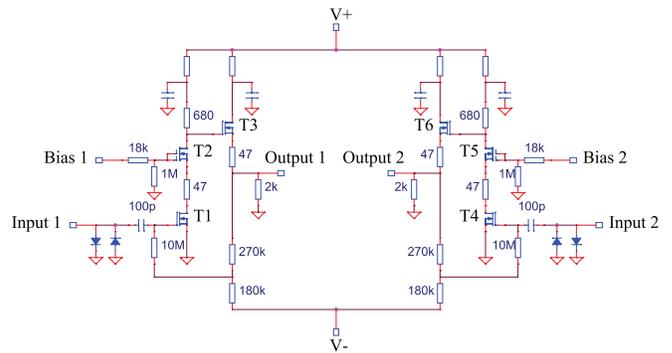}
\caption{Schematic of the cryogenic  low-noise amplifier stage. This design gives a broadband frequency response in the low-MHz  regime. Both channels are mounted on the same printed circuit board. The amplifier settings can be tuned by the source follower biasing for both channels simultaneously, and through  the cascode biasing for each channel independently.  \label{demo_HF_cryo}}
\end{figure}

The cryogenic amplifiers are based on cascode biasing to the source follower. The cascode stage eliminates the Miller effect, which is the drain to gate capacitive feedback. The Miller effect also decreases the stability of the circuit at higher frequencies, where effective feedback is positive due to the large phase shift ($>180^\circ$) between the input and the feedback signal.  This feedback limits the upper cutoff frequency in the source follower circuit. The source follower is a high input impedance amplifier and the output impedance of this stage matches the cable impedance hence improves the high frequency performance. 

The input capacitance of the two amplifiers combined, including the that due to the short cables of about 1~cm, amounts to a total of $C_T=14.3$~pF.  
Assuming the impedance $|Z_{in}|$ of the circuit at the input, which we take to include the stray capacitance $C_T$, is much smaller than the amplifier input resistance, the effective noise due to the amplifiers detected in channel 1 is given by (see Fig.~\ref{fig:circuit})
\begin{equation}
V^2_{n,1} = v_{n,1}^2 + i_{n,1}^2|Z_{in}|^2 + i_{n,2}^2|Z_{in}|^2 +c_n^2V_{n,2}^2 .   \label{eq:residual_noise}
\end{equation}
Here $v_{n,i}$ and $i_{n,i}$ are the voltage and current noise, respectively, at the input of the amplifier number $i$, while the $c_n$ is a factor describing the cross talk. The voltage noise of the first-stage amplifiers is 0.48~nV/$\sqrt{\rm Hz}$. The current noise of 60~fA/$\sqrt{\rm Hz}$ would normally dominate over the voltage noise for resistor values in the range of the resistance quantum. However, the current noise is suppressed by the instrumental bandwidth $|Z_{in}(\omega)|^2$ along with the sample noise signal, which does not apply for the voltage noise. 
The voltage noise of the two amplifiers, on the other hand, can be nearly fully suppressed by the cross spectrum technique, because they are uncorrelated. Both channels for the cryogenic amplifier are on same board, which gives rise to a small cross talk $c_n$ of about $1\%$ of the full signal at each stage. Although this limits the effectiveness of the cross spectrum and could be further improved by mounting the two channels on separate boards, in practice this is not a limiting factor.

The gain for each of the cryogenic amplifiers of 7.5$\times$ 
is high enough to make the input noise of 0.25~nV/$\sqrt{\rm Hz}$ of the room temperature amplifier stages negligible. The gain of the amplifiers is adjustable through the supply settings and the gain of two channels can be made to match to within  less than 1dB. 
The room-temperature amplifiers have a gain of 200 and are mounted directly at the brass head on the top of the dipstick. The output lines are coupled to the input of the PXI data acquisition system through 50~$\Omega$ coaxial cables. 

\subsection*{Digital electronics and data manipulation.}

We have adopted a National Instruments data acquisition unit for the digitization of the signals. It comprises the NI 5762(-01) with two simultaneously sampled 16-bit analog-to-digital convertor (ADC) channels at 2~V range, 250~MS/s and 100~MHz bandwidth. The synchronization of the sampling by the two channels is critical for obtaining an undistorted cross spectrum. The channels are ac coupled and equipped with a 100~MHz elliptic filter. 
It has a low $-110$dBFS channel-to-channel crosstalk. 
The  ADC adds noise of about 10~nV/$\sqrt{\mathrm{Hz}}$ referred to the input of the ADC channels 
which is negligible in view of the amplification level before the ADC. 

The data handling speed is accelerated by using a NI FlexRIO FPGA module (NI PXI-e-7966R). Due to the $32$-bit PXI express bus with direct memory access capability data communication from the FPGA to the host computer and to the FPGA from the host computer within the  PXI backplane can be done above $100$MB/s and $50$MB/s, respectively. The combination of the PXI express bus and FPGA allows the implementation of a real time cross spectrum analyzer. The FPGA is programmed for handling the consecutive blocks of 2048 digitized data points from both channels, for applying a Hann windowing function, performing fast Fourier transform (FFT) of both channels individually, computing the complex cross spectrum, vector averaging by an adjustable number of up to 10$^6$ spectra, and computing the power spectrum of the result. The FFT processing time for blocks of 2048 data points is 4$\mu$s, implying that the computation time is less than the data acquisition time, making the latter the limiting factor in the speed of measurement. The final averaged power spectrum is transferred to a desktop PC for further processing and storage.

\begin{figure}[!h]
\includegraphics[scale =0.6]{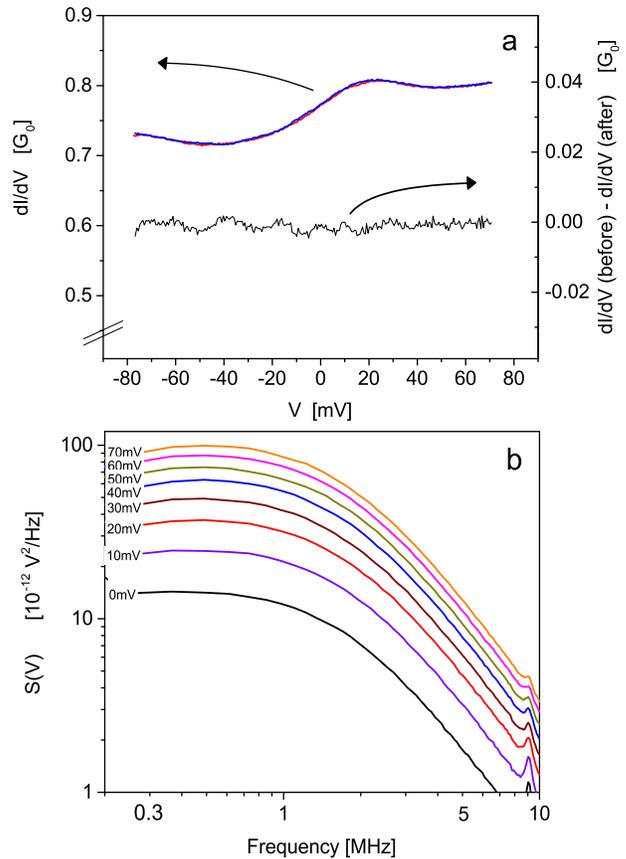}
\caption{Example of raw measurement data for a Au atomic contact at 4.2~K. (a) Differential conductance characterizing the junction configuration before (red) and after (blue) recording the sequence of noise spectra. In this case the red and blue curves nearly coincide, and the difference between the curves is displayed by the black curve (right axis). The zero-bias conductance is $G(0) = 0.772\,G_0$. (b) Noise spectra for the same junction setting at various levels of applied dc bias voltage.    \label{fig.raw_data}}
\end{figure}

The measurements are controlled from a dedicated user-built Labview routine running on a desktop computer. A typical measurement procedure proceeds along the following steps. The routine automatically controls the breaking of an atomic junction through control of the piezo voltage, and searches for a target conductance value. Having arrived at a desired junction setting the bending force on the sample substrate is held stationary. The program then switches to taking a differential conductance spectrum by sweeping the bias voltage between user-defined limits, typically from $-80$~mV to $+80$~mV. After the dc bias has been returned to zero, a first noise spectrum is recorded, which is the thermal noise signal $S_m(0)$, against which the noise at finite bias is compared. Next, shot noise spectra $S_m(V)$ are recorded at user-defined steps of typically 0.5~mV, up to a user-defined maximum, typically 80~mV. This series of spectra can be recorded in about 14 seconds. In order to test for changes in the atomic contact configuration during the measurements a differential conductance spectrum is taken again for the same junction setting.
All raw data are stored for later inspection, but the analysis is done in real time by a parallelly operating routine.  Figure~\ref{fig.raw_data} shows an example of a set of raw data. The small peak at 9~MHz is due to a signal picked up from the piezo-voltage source, but it does not affect our results, see below.

For the analysis of the Fano factor we would like to make use of relation (\ref{eq.YvsX}), which has the advantage that it is linearly dependent on the quantity $X(V,T)$ and deviations from a linear dependence can be easily detected. However, this is only strictly valid on the condition that the measured zero-bias noise is purely the thermal noise of the junction under test. We have identified at least two additional external noise sources that need to be taken into account: the thermal noise of resistor $R_1$ in the bias line (see Fig.~\ref{fig:circuit}), and the current noise $i_n$ injected by the input of the amplifiers into the junction circuit. When $S(V)$ describes the true noise power spectral density due to the junction, the measured noise power can be written as,
\begin{equation}
S_m = |\beta(\omega)|^2 \left[ S(V) |H_1(\omega)|^2 + \tilde{S}_{\mathrm{x}}(T,\omega)\right] . \label{eq.Sm}
\end{equation}
The intrinsic and extrinsic noise sources can be added independently because they are uncorrelated. Here, $H_1(\omega)$ is the transfer function of the noise signal due to the junction, as seen at the input of the amplifiers. It is described in terms of the capacitors and resistors in the input circuit, including the junction resistance $R_s$ and input capacitance $C_T$: $H_1 = Z/(R_s+Z)$. Here, $Z$ is the impedance of the input capacitance, $1/j\omega C_T$ in parallel with $R_1 + Z_\mathrm{ext}$, where $Z_\mathrm{ext}$ represents the impedance of the circuit beyond resistor $R_1$. In (\ref{eq.Sm}) the term $\tilde{S}_{\mathrm{x}}(T,\omega)$ represents all noise sources other than that due to the atomic junction, and the frequency dependence of the transfer function is incorporated in its definition. Finally $\beta(\omega)$ is the transfer function of the amplifier chain, including the amplification factor.

\begin{figure}[!h]
\includegraphics[scale =0.6]{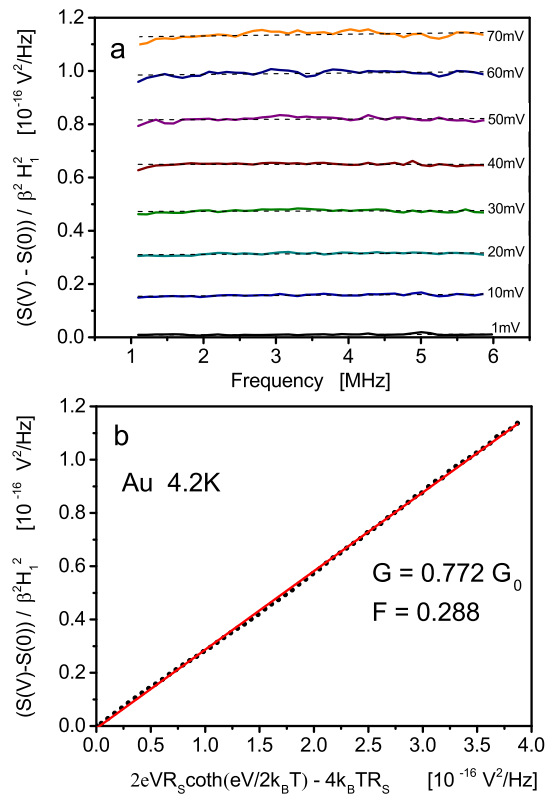}
\caption{ Determination of the Fano factor from shot noise measured on a Au atomic contact at 4.2~K. (a) The excess noise $S(V)-S(0)$  obtained from the measured data in Fig.~\protect\ref{fig.raw_data}, corrected for amplifier and transfer characteristics, shown for selected bias voltages as a function frequency. The mean value for each curve in  is obtained from  linear fits (shown in red), and the resulting values are plotted in panel (b) as a function of the scaled bias voltage $4k_\mathrm{B}TR_\mathrm{s}(X(V,T) - 1)$, which can also be written as $ 2eVR_\mathrm{s}\coth (eV/2k_{\rm B}T) - 4k_\mathrm{B}TR_\mathrm{s} $. Since the hyperbolic cotangent very quickly saturates at 1 the latter expression shows that the scale is nearly proportional to the applied bias voltage $V$, which in this case runs up to about 70~mV. The term $4k_\mathrm{B}TR_\mathrm{s} $ is only $0.038\cdot10^{-16}$~V$^2$/Hz. From the slope of this reduced axis plot we obtain a Fano factor of $F = 0.288\pm0.004$.
  \label{fig.reduced_noise}}
\end{figure}

By subtracting the zero-bias noise from the shot noise measured at finite bias, $S_m(V) - S_m(0)$, the external noise sources can be eliminated. From this difference we obtain the true excess noise $S(V) - S(0)$ by first dividing out the amplification characteristics $ |\beta(\omega)|^2 $ obtained from a calibration step described below, and dividing out the  transfer function $|H_1(\omega)|^2$ that we obtain from the known components of the input circuit. We verify that the resulting excess noise spectra given by
\begin{equation}
S_e = \frac{S_m(V) - S_m(0)}{|\beta(\omega) H_1(\omega)|^2 }          \label{eq.excess} 
\end{equation}
are white (i.e. frequency independent) in an appropriate frequency window, see Fig.~\ref{fig.reduced_noise}a. Linear fits to the spectra in Fig.~\ref{fig.reduced_noise}a measure the slope and the mean noise power amplitude. 

The resulting mean excess noise is expected to be equal to the function $Y(V,T)$, Eq.~(\ref{eq.YvsX}), multiplied with the nominal thermal noise of the junction $4k_{\mathrm{B}}TR_s$, for the measured value of the junction resistance $R_s$. Thus, we proceed to plot the excess noise (\ref{eq.excess}) against $4k_{\mathrm{B}}TR_s(X(V,T) - 1)$.  In this product form for $X$ the sample temperature is nearly perfectly eliminated from the analysis, since the $\coth$ function rapidly saturates at 1, and the term $-1$ is small compared to $X$. An example of the resulting plot is shown in Fig.~\ref{fig.reduced_noise}b, from which the Fano factor $F$ is determined as the slope of a linear fit to the data points.
 
At the end of the noise measurement and analysis for a given junction setting the program searches for a new junction in the desired conductance range and restarts the sequence. 
  
\section*{Testing, calibration, and accuracy}
 
The gain $\beta$ of the amplifiers and the input capacitance $C_T$ are characterized by injecting a known white noise signal from an  Agilent 33500B waveform generator through the bias line
for infinite sample resistance (broken junction). The amplitude of the noise signal needs to be selected carefully. The peak events in the noise need to stay below the saturation level of 2~V of the second-stage amplifiers. 
With a combined amplification of the two stages of 1500$\times$ the peak voltage at the input must remain below 0.67~mV, which corresponds to approximately 0.2~mV$_\mathrm{rms}$. Given the effective bandwidth of 1~MHz of the circuit this translates into a noise level of 100~nV$/\sqrt{\mathrm{Hz}}$. At the lower end the calibration signal is limited by the requirement that it needs to be much higher than the other noise sources in the circuit, such as the thermal noise of 1.5~nV$/\sqrt{\mathrm{Hz}}$ of resistor $R_1$. We use a signal level of 5~mV$_\mathrm{pp}$ terminated by 50~$\Omega$ at the bias line input,  behind resistor $R_0$. We verified that the gain $\beta$ obtained from this procedure is independent of the input amplitude  to within 1\%, over a range of a factor of 2 around this value. For larger and smaller amplitudes we find deviations, as expected.

The measured calibration noise power spectrum is closely reproduced by the circuit model parameters, with $\beta$ and $C_T$ as adjustable parameters. In good approximation the noise power shows a single corner frequency determined by the resistor $R_1$ and the input capacitance $C_T$. Since the amplification shows a small drift the calibration is done at the start and at the end of a measurement series, and occasionally also at intervals during a series.

The junction is expected to be at the bath temperature of 4.2K. However, due to heating of the amplifiers and the direct connection of short wires between the junction and the amplifiers a slight increase in sample temperature cannot be excluded. By fitting the measured noise $S_m(V,T)$ as a function of bias at low bias we obtain a measure of the true junction temperature as $4.4\pm0.2$K. 

Let us now consider the factors that influence the accuracy of the Fano factor. By the signal analysis procedure outlined in the previous section we eliminate the influence of many possible sources of inaccuracy: the uncertainty in amplifier current noise and voltage noise, cross talk of the amplifiers, uncertainty in the resistance $R_1$ and its noise temperature, and uncertainties in the junction temperature. We found that grounding of all electronics needs careful attention. Subtle ground loops are detected as a spurious rounding of the Fano plot at low applied bias, as a result of small ac induced currents at nominally zero bias. Many tests on Au atomic junctions confirm that we have effectively eliminated this problem, as illustrated in Fig.~\ref{fig.reduced_noise}b. The main uncertainty in the Fano factor derives from spontaneous (or current-induced) changes in the sample resistance during the noise measurements. This problem is addressed by comparing the differential conductance measured at the start and at the end of the cycle. We typically impose a limit of 5\% in difference between the two and discard all data with a larger change in conductance. The final accuracy in $F$ is limited by the calibration of the amplification $\beta$ (about 1\%) and by random statistics in the noise signal due to the finite averaging time, limiting the absolute accuracy to about 0.01.

The role of Joule heating in the sample and in the resistor $R_1$ was tested by placing a 10~k$\Omega$ resistor in the place of the sample and measuring the noise as a function of bias, as for atomic junctions. Since resistors have no shot noise any increase in noise detected can be attributed to Joule heating of the two resistors. It was necessary to increase the bias over the mock sample resistor up to 1~V in order to detect appreciable increase in noise. The noise increases approximately proportional to the square of the applied bias, as expected for Joule heating. This additional thermal noise remains well below $10^{-3}$ of the shot noise signals measured here, even up to 1~V in bias.

\section*{Measurements on Au and Pt atomic contacts}

\begin{figure}[!h]
\includegraphics[scale =1]{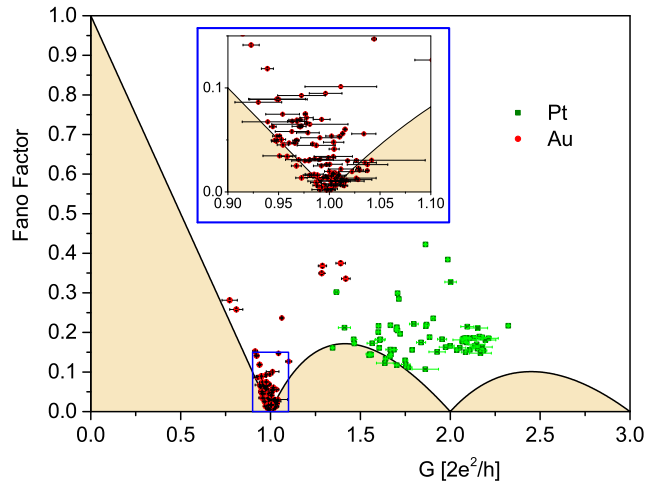}
\caption{ Fano factor plot for noise measurements on over $200$  atomic configurations for  Au and Pt. The Fano factors for all atomic contacts are above the theoretical limit of minimum Fano factor for a spin-degenerate Landauer conductor (full curve), within the accuracy of the measurements. The data cluster close to the boundary, indicating that a single partially transmitting channel frequently forms for such atomic contacts.} \label{fig.FanoPlot}
\end{figure}

The performance of the system was tested on the known noise properties of Au and Pt atomic contacts.\cite{brom99,vardimon13} For many
different junction settings noise spectra were recorded in a window from 122~kHz to 10~MHz, and for voltage bias values between 0 and 80~mV. 
An example of such series of spectra is given  in Fig.~\ref{fig.raw_data}(b) above, and the data was analyzed in terms of the excess noise and Fano factors as outlined above. 
The Fano factors obtained for the low-bias regime for over 200
Au and Pt atomic contacts are plotted against the conductance of the junctions in Fig.~\ref{fig.FanoPlot}.

The solid black curve shows the absolute minimum for Fano factors  for spin-degenerate conductance channels. The curve forms the boundary for the shaded region below, that is inaccessible for non-magnetic junctions. For conductances at or below 1~$G_0$ the data for Au cluster close to the black curve, which is the expected dependence for a single conductance channel. Also, we observe a strong clustering of data near the sharp minimum in the Fano factor for conductance near 1~$G_0$. This is in agreement with the known properties for Au atomic junctions, which have a high probability to show a conductance near 1~$G_0$, and have a single conductance channel.\cite{scheer98,brom99,agrait03}. The set of experimental data for Au form the most direct test of the experimental technique, and demonstrates the accuracy in $G$ and $F$.

For comparison we also show data recorded for Pt atomic contacts. The data are much more scattered and are found at higher conductance, typically around 1.5~$G_0$, and higher Fano factors. Through participation of the d-orbitals Pt atomic contacts have access to five conductance channels, which may even have some spin splitting.\cite{strigl15} In agreement with previous measurements in the frequency window below 100~kHz\cite{kumar14} we find that the all data fall outside the region that is inaccessible for spin-degenerate systems, and there is some clustering at the boundary.

\begin{figure}[!h]
\includegraphics[scale =0.4]{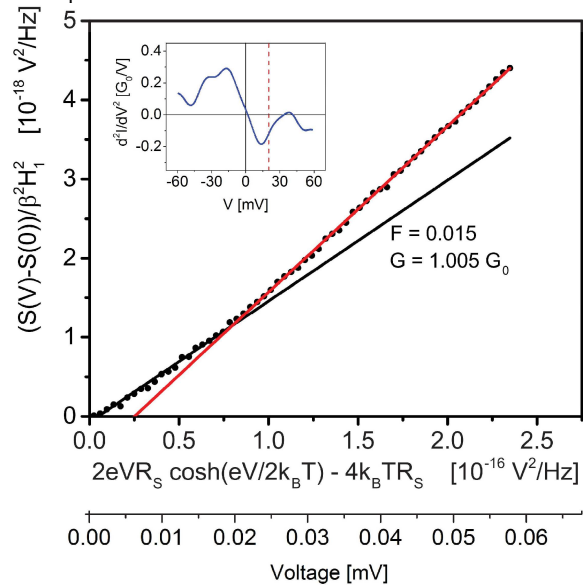}
\caption{Excess noise as a function of the scaled bias voltage for a Au atomic chain, showing a characteristic kink attributed to inelastic scattering. The position of the kink lies close to the position of the vibration mode energy seen in the second derivative spectrum for the current, shown by the red broken line in the inset. The bottom scale in the main figure gives an approximate scale for the corresponding bias voltage.} \label{fig.kink}
\end{figure}

Figure~\ref{fig.reduced_noise} illustrates the fact that the present technique permits accurate measurements of shot noise to very high bias values. For the example given in the figure the inelastic scattering is weak (as also evidenced by the differential conductance spectrum in Fig.~\ref{fig.raw_data}a) and we find that the excess noise is linear with the scaled bias parameter up to more than 70~mV. Interesting deviations from this purely ballistic electron transport are found for junctions with a stronger electron-vibron scattering amplitude. An example for Au atomic chains is given in Fig.~\ref{fig.kink}, which shows a clear kink in the noise plot near the point where the peak in $d^2 I/dV^2$ indicates the presence of a vibron excitation. Such inelastic scattering effects have been investigated in the low-bias regime for Au atomic chains in previous work,\cite{kumar12} which can now be extended by the present technique to molecular junctions with higher vibron mode energies. 

\begin{figure}[!h]
\includegraphics[scale = 1]{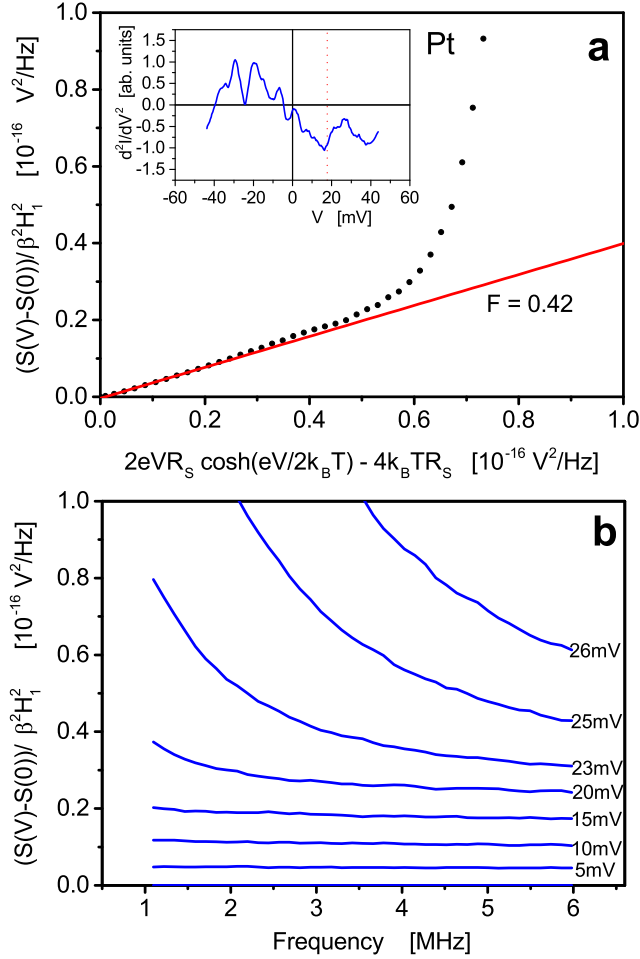}
\caption{Example of a junction showing anomalous noise. (a) At low bias, up to about 7~mV, the excess noise is perfectly linear giving a Fano factor $F=0.42$. At about 20~mV ($0.6\cdot10^{-16}$ in the reduced units) a very strong deviation develops. The differential conductance spectrum in the inset is suggestive of a vibron mode at this energy. (b) However, the freqency spectrum clearly shows that the noise is not white, and a very pronounced frequency dependence develops at higher bias.} \label{fig.run-away}
\end{figure}

The benefits of having access to the spectral information is further illustrated by the example for a Pt junction in Fig.~\ref{fig.run-away}. For bias voltage settings up to about 10~mV the noise remains white and the excess noise rises linearly with the scaled bias parameter, as expected. For higher bias the junction develops anomalous behavior, with giant deviations from the linear excess noise dependence. At the same time the appearance of the noise spectrum changes from the usual white spectrum into a spectrum with a pronounced rise towards lower frequencies. This spectral information assists in analyzing the source of this noise, which obviously is not just shot noise. Such spectral shapes require processes with time scales that are long compared to typical atomic or electronic processes. We tentatively attribute the noise to two-level fluctuations, and a more definite interpretation is deferred to future publications.

\section*{Conclusions}

The benchmark method for shot noise measurements on atomic contacts employs the cross spectrum of the signal from two sets of amplifiers in a frequency window up to 100~kHz. By introducing cryogenic amplifiers close to the atomic junction the input capacitance was reduced by nearly two orders of magnitude. The resulting wider frequency band allows us to measure spectra up to10~MHz, which is necessary when 1/f noise is likely to interfere with the shot noise signal, as expected to occur for molecular junctions and at higher voltage bias. Working at higher frequencies has the additional advantage that data acquisition and averaging of the spectra can be much faster. In order to profit from this advantage, the Fourier transform and other spectral manipulation is programmed on an FPGA module, such that the speed of measurement is only limited by data acquisition. The acquisition time scales with the inverse of the lowest frequency in the spectral window, and by shifting the spectral window for Fourier transform from 250~Hz--100~kHz to a window of 122~kHz--100~MHz the measurement time is reduced by a factor 500. The standard technique requires nearly a minute of signal averaging for each measurement of noise spectrum for a given junction and a given bias voltage. Since the Fano factor is obtained from the plot of noise for many bias setting, the measurement of the Fano factor for a junction takes many minutes. Detecting subtle deviations in the linear dependence of noise with bias\cite{kumar12} requires many bias points, and the total measurement time per junction setting may be as long as 30 minutes. The wide band technique introduced here reduces this measurement time to 3 -- 4 seconds. This method may also find application in scanning tunneling microscopy experiments.\cite{burtzlaff15}

Further improvements are possible. By redesign of the amplifier circuit board a further reduction of the input capacitance by a factor of three should be possible. However, a more significant increase in frequency can be obtained by adding an inductor in parallel to the junction.\cite{dicarlo06} For example, the present low-pass characteristics with a corner frequency of 1.5~MHz can be converted into a band-pass characteristics with a center frequency of 40~MHz for an inductor of 1~$\mu$H. This will be attractive when 1/f noise is still detectable near 1~MHz, and it will also permit further increasing the speed of measurement.

\section*{ACKNOWLEDGMENTS}

This work was supported by the Netherlands Organization for Scientific Research (NWO/OCW), as
part of the Frontiers of Nanoscience program and is part of the research programme of the Foundation for Fundamental Research on Matter (FOM),
which is financially supported by NWO. We thank  A. van Amersfoort and R. van Egmond for expert technical support, and T.A. de Jong for assistance in the experiments.

\bibliographystyle{utphys}
\bibliography{RevSciInstrum}

\end{document}